\documentclass[aps,prl,twocolumn,superscriptaddress]{revtex4-1}

\usepackage{graphicx}
\usepackage{color}
\usepackage{here}

\begin{document}

%\preprint{}

\title{Evolution of remnant Fermi surface state in lightly-doped correlated spin-orbit insulator Sr$_{2-x}$La$_x$IrO$_4$}


\author{K. Terashima}
\email[]{k-terashima@cc.okayama-u.ac.jp}
%\thanks{}
\affiliation{Research Institute for Interdisciplinary Science, Okayama University, Okayama, 700-8530, Japan}

\author{M. Sunagawa}
\affiliation{Graduate School of Natural Sciences, Okayama University, Okayama, 700-8530, Japan}
\author{H. Fujiwara}
\affiliation{Graduate School of Natural Sciences, Okayama University, Okayama, 700-8530, Japan}
\author{T. Fukura}
\affiliation{Graduate School of Natural Sciences, Okayama University, Okayama, 700-8530, Japan}
\author{M. Fujii}
\affiliation{Graduate School of Natural Sciences, Okayama University, Okayama, 700-8530, Japan}
\author{K. Okada}
\affiliation{Aoyama Gakuin University, Sagamihara, kanagawa 229-8558, Japan}
\author{K. Horigane}
\affiliation{Research Institute for Interdisciplinary Science, Okayama University, Okayama, 700-8530, Japan}
\author{K. Kobayashi}
\affiliation{Research Institute for Interdisciplinary Science, Okayama University, Okayama, 700-8530, Japan}
\affiliation{Graduate School of Natural Sciences, Okayama University, Okayama, 700-8530, Japan}
\author{R. Horie}
\affiliation{Research Institute for Interdisciplinary Science, Okayama University, Okayama, 700-8530, Japan}
\author{J. Akimitsu}
\affiliation{Research Institute for Interdisciplinary Science, Okayama University, Okayama, 700-8530, Japan}
\author{E. Golias}
\affiliation{Helmholtz-Zentrum Berlin f$\ddot{u}$r Materialien und Energie, Albert-Einstein-Str. 15, 12489 Berlin, Germany}
\author{D. Marchenko}
\affiliation{Helmholtz-Zentrum Berlin f$\ddot{u}$r Materialien und Energie, Albert-Einstein-Str. 15, 12489 Berlin, Germany}
\author{A. Varykhalov}
\affiliation{Helmholtz-Zentrum Berlin f$\ddot{u}$r Materialien und Energie, Albert-Einstein-Str. 15, 12489 Berlin, Germany}
\author{N. L. Saini}
\affiliation{Dipartimento di Fisica, Universit\'{a} di Roma ``La Sapienza'' - P. le Aldo Moro 2, 00185, Roma, Italy }
\author{T. Wakita}
\affiliation{Research Institute for Interdisciplinary Science, Okayama University, Okayama, 700-8530, Japan}
\author{Y. Muraoka}
\affiliation{Research Institute for Interdisciplinary Science, Okayama University, Okayama, 700-8530, Japan}
\affiliation{Graduate School of Natural Sciences, Okayama University, Okayama, 700-8530, Japan}
\author{T. Yokoya}
\affiliation{Research Institute for Interdisciplinary Science, Okayama University, Okayama, 700-8530, Japan}
\affiliation{Graduate School of Natural Sciences, Okayama University, Okayama, 700-8530, Japan}


\date{\today}

\begin{abstract}
Electronic structure has been studied in lightly electron doped correlated spin-orbit insulator Sr$_2$IrO$_4$ by angle-resolved photoelectron spectroscopy.  We have observed coexistence of the lower Hubbard band and the in-gap band, the momentum dependence of the latter traces that of the band calculations without on-site Coulomb repulsion.  The in-gap state remained anisotropically gapped in all observed momentum area, forming a remnant Fermi surface state, evolving towards the Fermi energy by carrier doping.  These experimental results show a striking similarity with those observed in deeply underdoped cuprates, suggesting the common nature of the nodal liquid states observed in both compounds.
\end{abstract}

%\keywords{}

\maketitle
	Unconventional physics of superconductivity near the metal-insulator transition in strongly-correlated Mott insulators has been one of the major themes in a variety of systems such as cuprates, iron-based compounds, heavy-electron systems, and organic materials\cite{RMP}.  Recently, 5{\it d}-electron systems are gaining large attention in which the magnitude of spin-orbit coupling is comparable to transfer-integral and Coulomb repulsion energies, and this interplay may produce possible novel phases.  Sr$_2$IrO$_4$ is a good example of such system for which the electronic states can be well described by considering spin-orbit coupling as well as Coulomb repulsion energy {\it U} \cite{KimPRL, RIXS}.
	
	Sr$_2$IrO$_4$ is an antiferromagnetic insulator with {\it T}$_{\rm N}$ = 240 K, and is isostructural to one of the parent compound of cuprate superconductors namely La$_2$CuO$_4$\cite{Crawford}.   Similar to the cuprates, the electronic structure is highly two-dimensional, revealed by angle-resolved photoelectron spectroscopy (ARPES)\cite{QWang, Yamasaki}. Unlike cuprates, to date Sr$_2$IrO$_4$ does not show superconductivity although a possible emergence of superconductivity in this system has been theoretically predicted by carrier-doping\cite{FWang,Watanabe,YYang,ZYMeng}.  On the other hand, {\it d}-wave gapped state and the Fermi arc behavior has been observed in both bulk\cite{Delatorre} and surface\cite{KimScience,KimNP,YJYan} electronic structure of doped Sr$_2$IrO$_4$, similar to the cuprates. Such a similarity is puzzling and raises several questions merely due to lack of momentum-resolved data in a wide range of doping especially in deeply underdoped regime. This is indeed crucial to explore if this anisotropic gap has the same origin as pseudogap in cuprate superconductors and if the gap is related to the superconductivity.

\begin{figure}[b]
\includegraphics{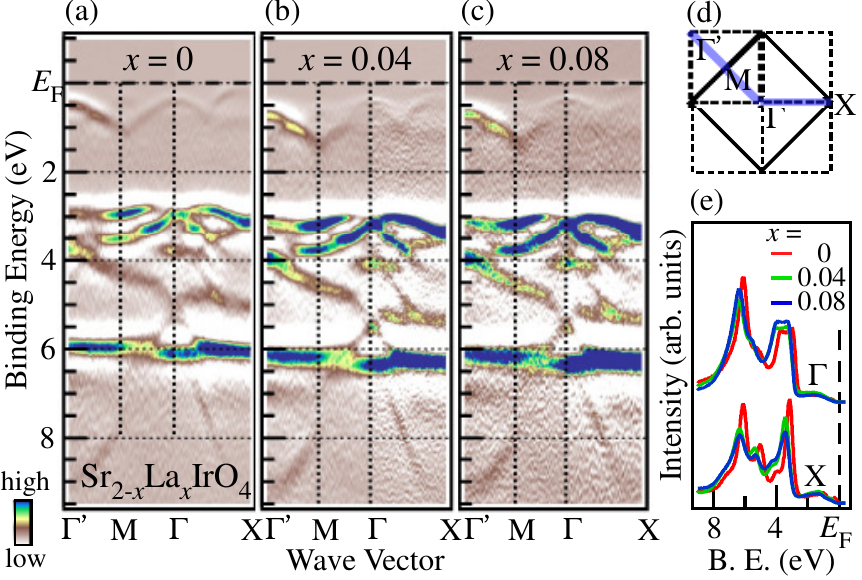}%
\caption{(color online). (a)-(c) Second derivative of ARPES intensity plots as a function of binding energy and wave vector on Sr$_{2-x}$La$_x$IrO$_4$ ({\it x} = 0, 0.04, and 0.08) measured along {\it k}-direction shown as a blue line in (d).  (d) Solid (dashed) line shows the folded (unfolded) Brillouin zone. (e) EDCs at the $\Gamma$ and the X points for {\it x} = 0, 0.04, and 0.08.}
\end{figure}	
	
\begin{figure*}[t]
\includegraphics{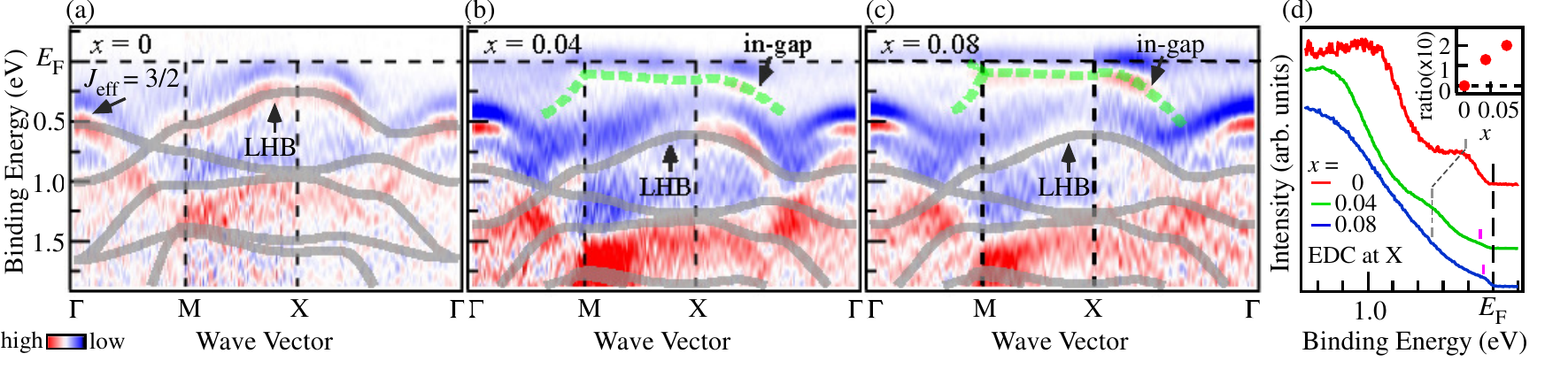}%
\caption{(color online). (a)-(c) Second derivative of ARPES intensity plots in the near-{\it E}$_{\rm F}$ region along high symmetry line on {\it x} = 0, 0.04, and 0.08, respectively.  Overlaid gray curves in (a) are calculated band by tight-binding method\cite{Delatorre} where on-site coulomb repulsion energy is 2 eV.  The same curves are shown in (b) and (c) after shifting downward by 0.3 eV, according to the estimated energy shift of entire valence band in Fig. 1.  Green dashed curves in (b) and (c) are the guide for in-gap state (see also Figs. S2 and S3). (d) EDCs at X point for different samples.  Gray bars denote the position of LHB, and purple bars denote the position of in-gap state.  Inset shows the relative intensity ratio at peak/hump positions $\sim$ 0.5 eV and 0.1 eV as a function of doping, which depicts the relative weight of LHB and in-gap state.  The positions were chosen according to the peak positions in the second derivative of EDCs.}
\end{figure*}

	To address these unsettled issues, we have studied how this {\it d}-wave gapped state evolves by doping in lightly-doped Sr$_{2-x}$La$_x$IrO$_4$ ({\it x} = 0, 0.04, 0.08) using ARPES.  We have observed a dispersive in-gap state that evolves by carrier doping and coexists with the lower Hubbard band (LHB) seen in the parent compound\cite{KimPRL}. The in-gap state appears gapped in all observed momentum ({\it k})-region with anisotropy in the gap magnitude. This behavior is remarkably similar to the remnant Fermi surface state observed in deeply underdoped cuprates\cite{Ronning}.  The results imply that the nodal liquid state\cite{Delatorre} with {\it d}-wave gap symmetry in doped Sr$_{2-x}$La$_x$IrO$_4$ may hold a common nature to the one in cuprate superconductors.

	Single crystals of Sr$_{2-x}$La$_x$IrO$_4$ ({\it x} = 0, 0.04, 0.08) were synthesized by the flux method\cite{Sung}, and were used for ARPES measurement. La content of samples were determined by electron-probe microanalysis.  The temperature dependence of resistivity and magnetic susceptivility of samples are shown in supplemental information (Fig. S1).
	
	ARPES experiments were performed at 1-squared beamline of BESSY II, using Scienta-Omicron R8000 analyzer.  Circularly polarized light with {\it h}$\nu$ = 100 eV was used to excite photoelectrons.  Clean surfaces for measurements were obtained by {\it in situ} cleaving of samples, and they were measured in an ultrahigh vacuum better than 1$\times$10$^{-10}$ Torr.  Sample surfaces did not show any sign of degradation during measurements.  Fermi level ({\it E}$_{\rm F}$) of samples were estimated by that of copper plate electronically contacted with samples.  The energy- and angular-resolutions were set at 20 meV and 0.5 degee (corresponding to $\sim$0.04 $\rm {\AA}$$^{-1}$), respectively. The data were taken at  {\it T} = 100 K for {\it x} = 0 to avoid possible charging effect\cite{KimPRL}, and  {\it T} = 40 K for {\it x} = 0.04 and 0.08 samples, except for the {\it T}-dependence data in Fig. 4.

	Figures 1 (a)-(c) show second derivative of ARPES intensity as a function of wave vector and binding energy for Sr$_{2-x}$La$_x$IrO$_4$ ({\it x} = 0, 0.04, and 0.08). The color bar is shown to represent the intensity scale.  The data were taken along high symmetry line (blue line) shown in Fig. 1(d), where solid (dashed) lines denote the Brillouin zone with (without) the rotation of IrO$_6$ octahedra.  Energy distribution curves (EDCs) at representative momenta are shown in Fig. 1(e).  We have observed a number of dispersive bands up to binding energy $\sim$ 9 eV in {\it x} = 0 sample.  Almost identical bands are seen in {\it x} = 0.04 sample, but their energy positions are shifted downwards by $\sim$0.3 eV at every measured momenta, showing rigid-shift like behavior.  On the other hand, the energy bands in {\it x} = 0.08 sample are at very close to those of {\it x} = 0.04 sample, indicating that doped carriers for {\it x}$<$0.04 shift the chemical potential and then the chemical potential is pinned against further doping up to {\it x} = 0.08.  Such a behavior in the valence band region is consistent with earlier observation in near-{\it E}$_{\rm F}$ region\cite{Brouet}.

	Figures 2 (a)-(c) depict the near-{\it E}$_{\rm F}$ region of the band dispersions for {\it x} = 0, 0.04, and 0.08.  See also Fig. S2 where raw spectral intensity and EDCs are displayed for each momentum cuts.  Overlapped gray curves are bands\cite{Delatorre} calculated by tight-binding method, in which both spin-orbit coupling and on-site Coulomb repulsion ({\it U} = 2 eV) were taken into account.  The overall character of near-{\it E}$_{\rm F}$ band dispersion in {\it x} = 0 sample has been well described by first principle calculations considering both spin-orbit coupling and {\it U} and ascribed as {\it J}${\rm_{eff}}$ = 3/2 band and LHB of {\it J}${\rm_{eff}}$ = 1/2 band\cite{KimPRL, QWang, Yamasaki, Delatorre,Brouet, Moser}. In {\it x} = 0.04 and 0.08 samples, the corresponding bands (guided by gray curves) are shifted downwards by $\sim$0.3 eV as in the valence band region.  In addition to this rigid band behavior, we have found that additional dispersive states evolve in the vicinity of {\it E}$_{\rm F}$ by electron-doping. For example, a downward dispersion topped at the $\Gamma$ point in $\sim$0.5 eV and a flat band in MX direction at $\sim$0.1 eV that moves downward in X$\Gamma$ direction (shown as green dashed curves, see also Figs. S2 and S3).  With increasing carrier number, these bands become more clear while the LHB is getting smeared out.  This tendency is also visible in EDCs of {\it x} = 0.04 and 0.08 samples at X point shown in Fig. 2(d), where spectral intensity of LHB at $\sim$0.5 eV (gray bar) decreases while that of in-gap state at lower binding energy (purple bar) increases by electron-doping (see also Fig. S2).  Inset of Fig. 2(d) shows the intensity ratio between LHB and near-{\it E}$_{\rm F}$ structure shown by bars in Fig. 2(d). 
	
	The present data shows that two types of dispersive states coexist in lightly-doped region and doped carriers seem responsible for moving the spectral weight from LHB at high binding energy to in-gap state at low binding energy. It should be mentioned that the concept of the weight transfer has been already used in earlier ARPES study of EDCs at high symmetry points\cite{Brouet}. Here we show it explicitly revealing that both states exist as dispersive bands.  Such a coexistence of two states has not been observed in earlier ARPES study on {\it x} = 0.1 sample\cite{Delatorre}. This indicates that the spectral weight in LHB might have disappeared with a small additional electron doping than the present {\it x} = 0.08 sample.

\begin{figure}
\includegraphics{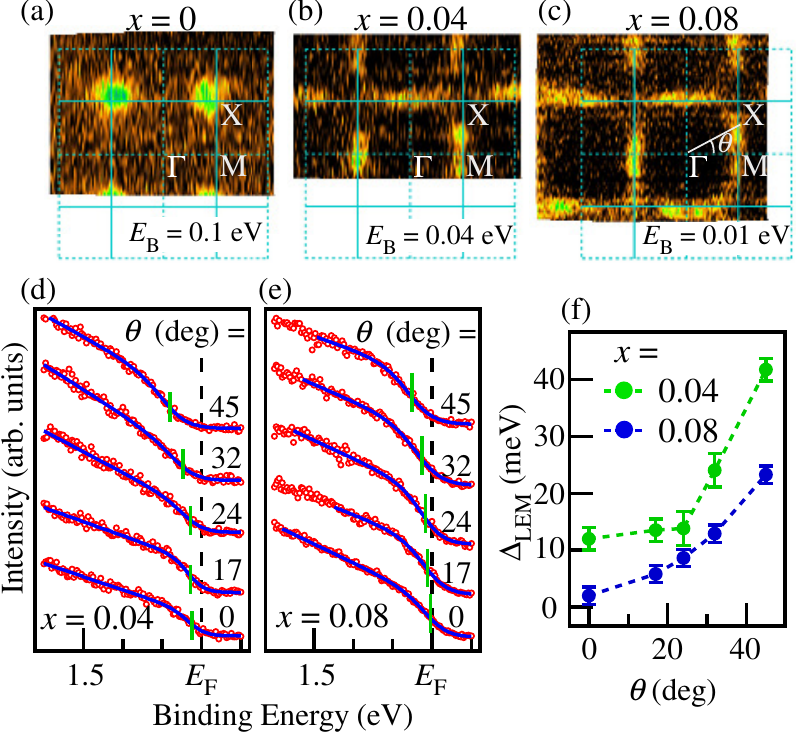}%
\caption{(color online). (a)-(c) ARPES intensity plot as a function of two dimensional wave vectors for Sr$_{2-x}$La$_x$IrO$_4$ {\it x} = 0, 0.04, 0.08, respectively.  The plots were taken at the binding energies shown in each figures, with an energy-integration window of $\pm$0.02 eV.  (d) and (e) EDCs taken at various {\it k}-points along energy contour on directional angles ($\theta$) for {\it x} = 0.04 and 0.08, respectively.  Blue curves are fitting results using Fermi-Dirac function, and green bars show obtained positions of leading edge midpoint.  (f) Estimated amount of energy shift in leading edge midpoint with respect to {\it E}$_{\rm F}$ ($\Delta$$_{\rm LEM}$) as a function of directional angle ($\theta$) for {¥\it x} = 0.04 (green) and 0.08 (blue).}
\end{figure}	
			
	To have further insight to the in-gap states, we have investigated the {\it k}-dependence in detail.  Figs. 3 (a)-(c) show ARPES intensity plots as a function of two-dimensional wave vector in Sr$_{2-x}$La$_x$IrO$_4$ at different binding energies that depict the energy contour at the top of the occupied state for each sample.  As in Fig. 3(a), the top of the LHB in parent compound appears at the X point, while the momentum distribution of in-gap state in {\it x} = 0.04 and 0.08 samples resembles to that in nodal liquid state reported in a sample with higher electron doping ({\it x} = 0.10)\cite{Delatorre}.  In Figs. 3(d) and (e), we have plotted EDCs extracted for various {\it k}-points along this energy contour. Corresponding positions of the {\it k}-points are defined by directional angle ($\theta$) in Fig. 3(c).  Blue curves in the figures are fitted results using Fermi-Dirac function where we set the chemical potential term as a free parameter\cite{Cao}. The leading-edge midpoint positions obtained by the fit are shown as green bars in the figure.  The magnitude of the shift of leading-edge midpoint with respect to {\it E}$_{\rm F}$ has been often used to evaluate the magnitude of gap\cite{Cao, Feng}.  It is clear from the figures that in-gap states are gapped and the magnitude of the gap is anisotropic.  In Fig. 3(f), we have shown the energy gaps estimated by leading-edge midpoint as a function of directional angle ($\theta$) which corresponds to {\it k}$_{\rm F}$-positions.  In both constituents, the gap values tend to be the smallest along $\Gamma$M direction ($\theta$ = 0 deg) and the largest along $\Gamma$X direction ($\theta$ = 45 deg).  With increasing doped electrons, the gapped state approaches {\it E}$_{\rm F}$ with keeping its overall anistropy in the lightly-doped regime of Sr$_{2-x}$La$_x$IrO$_4$.

\begin{figure}[t]
\includegraphics[width=3.4in]{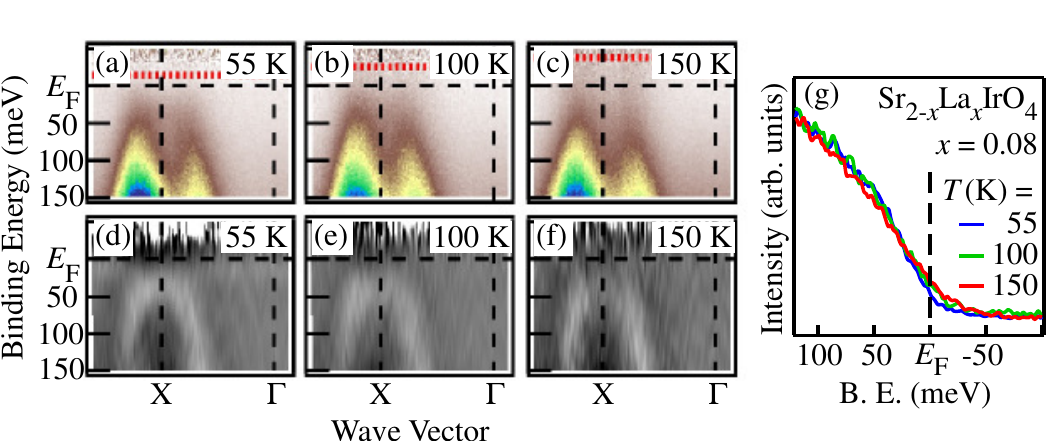}%
\caption{(color online). (a)-(c) Temperature dependence of ARPES spectra of {\it x} = 0.08 sample along $\Gamma$X direction divided by Fermi-Dirac function convoluted by experimental energy resolution, and their curvature plots ((d)-(f)). Red dashed lines in figures denote the energy corresponding to 3{\it k}$_B${\it T} above {\it E}$_{\rm F}$. (g) EDCs of (a)-(c) at estimated {\it k}$_{\rm F}$.
}
\end{figure}

	We show in Fig. 4 the temperature dependence of in-gap state in Sr$_{2-x}$La$_x$IrO$_4$ ({\it x} = 0.08) along $\Gamma$X direction, taken at {\it T} = 55, 100, and 150 K.  In order to obtain above-{\it E}$_{\rm F}$ region, the spectra were divided by Fermi-Dirac function convoluted by experimental energy resolution.  Figs. 4(d)-(f) are curvature plots\cite{curve} derived from (a)-(c).   We found that while the band is modified from that of LDA by the presence of gap at low temperature (see also Fig. S5(c)), the spectral weight of non-gapped band start to be filled by raising temperature. By this filling effect, the leading edge of {\it k}$_{\rm F}$ spectrum in Fig. 4(g) moves towards {\it E}$_{\rm F}$ as temperature raises but it is still away from {\it E}$_{\rm F}$ at {\it T} = 150 K, indicating that the gap has not yet completely closed at this temperature. We note that such a filling can be corresponding to the behavior of pseudogap in cuprates\cite{Reber}.

\begin{figure}[t]
\includegraphics{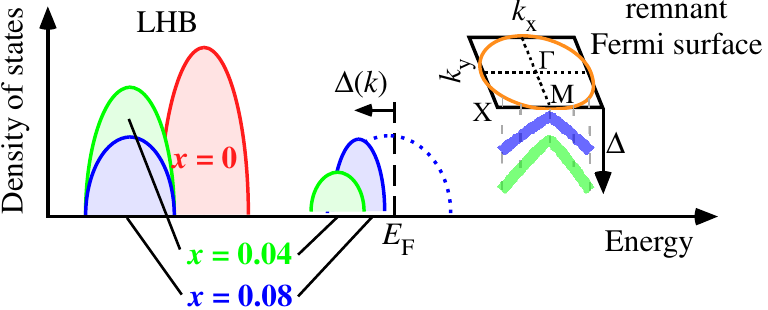}%
\caption{(color online). Schematic graph of doping-induced change in the electronic structure of Sr$_{2-x}$La$_x$IrO$_4$ derived from current study. Dashed blue curve represents tight-binding band expected to appear as in-gap state if there is no excitation gap.  Inset shows the remnant Fermi surface with gap anisotropy in the Brillouin zone.}
\end{figure}		
			
	Fig. 5 illustrates the schematic electronic structure deduced from above results.  As we dope electrons, the entire valence band including LHB shifts toward higher binding energy and pinned with formation of in-gap state.  Regarding the position of chemical potential with respect to the Mott gap, the position of LHB in doped samples ($\sim$0.5 eV) is close to the magnitude of Mott gap (0.54 eV) of parent compound observed by optical spectroscopy\cite{Moon}.  Thus the chemical potential is expected to lie at the tail of upper Hubbard band (UHB) as has been proposed earlier\cite{Brouet}, which would correspond to the case of (electron-) doped cuprates\cite{Steeneken}.  Further carriers are consumed to move the spectral weight of the LHB band to in-gap state below the Fermi level. Here we stress that this in-gap state is different from UHB, judging from the shape of the band dispersion. The in-gap state forms a dispersive band with momentum-dependence that tracks calculated tight-binding band but with anisotropic energy gap, resulting in a remnant Fermi surface (for more comparison of {\it E}-{\it k} dispersion with calculation, see Fig. S5).  The anisotropically gapped state approaches {\it E}$_{\rm F}$ by additional carrier doping as we observed at {\it x} = 0.08 sample.
	
	The observed remnant Fermi surface state shows a similar momentum dependence and gap anisotropy with the nodal liquid state reported in samples with higher electron doping\cite{Delatorre, KimNP}. Therefore the nodal liquid state seems to evolve from remnant Fermi surface state in electron-doped Sr$_2$IrO$_4$.  Our observation of coexistence of LHB and remnant Fermi surface state might correspond to a scanning tunneling microscopy results on Sr$_{2-x}$La$_x$IrO$_4$ where static phase separation between large gap state and small gap state has been observed\cite{Chen,Battisti}, while the present result is hardly reconciled with a simple phase separation of parent material and metallic state with fixed doping level since spectral lineshape and gapsize change by La doping in Fig. 3\cite{TNnote}. Our observation of momentum dependent band indicates that in-gap state would not be a completely localized state, therefore the phase separation would be, if exists, dynamical or static with coherence of finite distance.
	
	Next we compare our results of electron-doped system with ARPES results on Sr$_2$Rh$_{1-x}$Ir$_x$O$_4$\cite{Cao} where holes are effectively doped.  In the hole-doped system, it has been reported that LHB approaches {\it E}$_{\rm F}$ with keeping its band dispersion by hole doping.  This is contrary to the electron-doped system, where we observed the emergence of in-gap state whose momentum distribution is different from UHB.  This electron-hole asymmetry in the emergence of in-gap state upon carrier doping has been reproduced by first principles study\cite{PLiu}, except for the existence of anisotropic gap\cite{HWang} and remnant Fermi surface.  In the lightly hole-doped sample of Sr$_2$Ir$_{1-x}$Rh$_x$O$_4$ {\it x} = 0.04, LHB lies just below {\it E}$_{\rm F}$ with energy gaps in all {\it k}-region and with the anisotropy similar to higher hole-doping.  In this context, remnant Fermi surface seems to commonly exist in hole-doped and electron-doped system, although the gap anisotropy is reported to be Fermi-surface dependent in hole-doped system while it is rather momentum-dependent in electron-doped system.  Theoretical studies on such behavior on gap with electron-hole asymmetry would play a key role to clarify the nature of energy gaps observed in carrier-doped strontium iridate.
	
	Finally we discuss the correspondence/difference between Sr$_{2-x}$La$_x$IrO$_4$ and cuprates in the lightly-doped region.  First we note the correspondence in the existence of in-gap state. The coexistence of LHB and in-gap state has been observed in hole-doped La$_{2-x}$Sr$_x$CuO$_4$\cite{Ino} and electron-doped Nd$_{2-x}$Ce$_x$CuO$_4$\cite{Armitage, note} but not in other systems such as Ca$_{1-x}$Na$_x$CuO$_2$Cl$_2$\cite{Ronning} and Bi$_2$Sr$_2$CuO$_{6+\delta}$\cite{Zhou}. The remnant Fermi surface state has been observed in CaCuO$_2$Cl$_2$ where the band dispersion of the parent compound itself shows a {\it d}-wave gap anisotropy.  In case of Sr$_{2-x}$La$_x$IrO$_4$, the parent compound does not show such a {\it d}-wave gap anisotropy and the dispersion is well described by band calculation with spin-orbit interaction and {\it U}\cite{KimPRL}, while the in-gap state shows a behavior of remnant Fermi surface.  Second we mention the correspondence in the gap anisotropy as a function of doping. Despite these differences above, the remnant Fermi surface state shows a common feature as in cuprates in that the gapped states approach {\it E}$_{\rm F}$ by carrier doping with keeping their overall gap anisotropy\cite{Zhou}.  The present observation of remnant Fermi surface state suggests that Sr$_{2-x}$La$_x$IrO$_4$ system approaches to a metallic state by carrier doping in a highly corresponding manner with cuprate superconductors.

%\section{Summary}
	In summary, we have studied electronic structure of lightly electron-doped Sr$_{2-x}$La$_x$IrO$_4$.  We have found that a remnant Fermi surface is formed with anistropic gap, which approaches {\it E}$_{\rm F}$ and acquires spectral weight by increasing doped carriers.  These features show close similarity with those in cuprate superconductors, implying a common nature for nodal liquid states observed in both systems.
%\section{Acknowledgments}
	
	K. T. would like to thank Dr. K. Siemensmeyer for laue apparatus.  The authors are grateful to Mr. K. Tomimoto at Center for Instrumental Analysis for the measurements of elemental analyses.  K. T. and T. W. would like to acknowledge hospitality at the Sapienza University of Rome.  The experiments have been performed at 1-squared beamline of BESSYII.  This reseach was partially supported by the Program for Promoting the Enhancement of Research University from MEXT, the Program for Advancing Strategic International Networks to Accelerate the Circulation of Talented Researchers from JSPS, KAKENHI 25000003, and JSPS KAKENHI Grant Number 15H05886. The work is a part of the executive protocol of the general agreement for cooperation between the Sapienza University of Rome and the Okayama University, Japan.
\subsection{}
\subsubsection{}

\clearpage
\widetext
\begin{center}
\textbf{\large Supplemental Information}
\end{center}

\setcounter{figure}{0}
\setcounter{page}{1}
\renewcommand{\thefigure}{S\arabic{figure}}

\section{Sample properties}
	Figures S1(a) and S1(b) show the temperature dependence of the in-plane resistivity and the magnetic susceptibility for Sr$_{2-x}$La$_x$IrO$_4$ ({\it x} = 0, 0.04, 0.08), samples used for ARPES measurement.  The in-plane resistivity was measured in longitudinal resistivity configration by a standard four-probe method.  300 Oe of magnetic field parallel to ab-plane was applied for magnetic susceptibility measurements.  The $\chi$-{\it T} curve of parent compound is basically consistent with earlier report\cite{chi}.  Both of resistivity and magnetic susceptibility change systematically with increasing amount of La dopant.

\begin{figure}[h]
\includegraphics[width=6in]{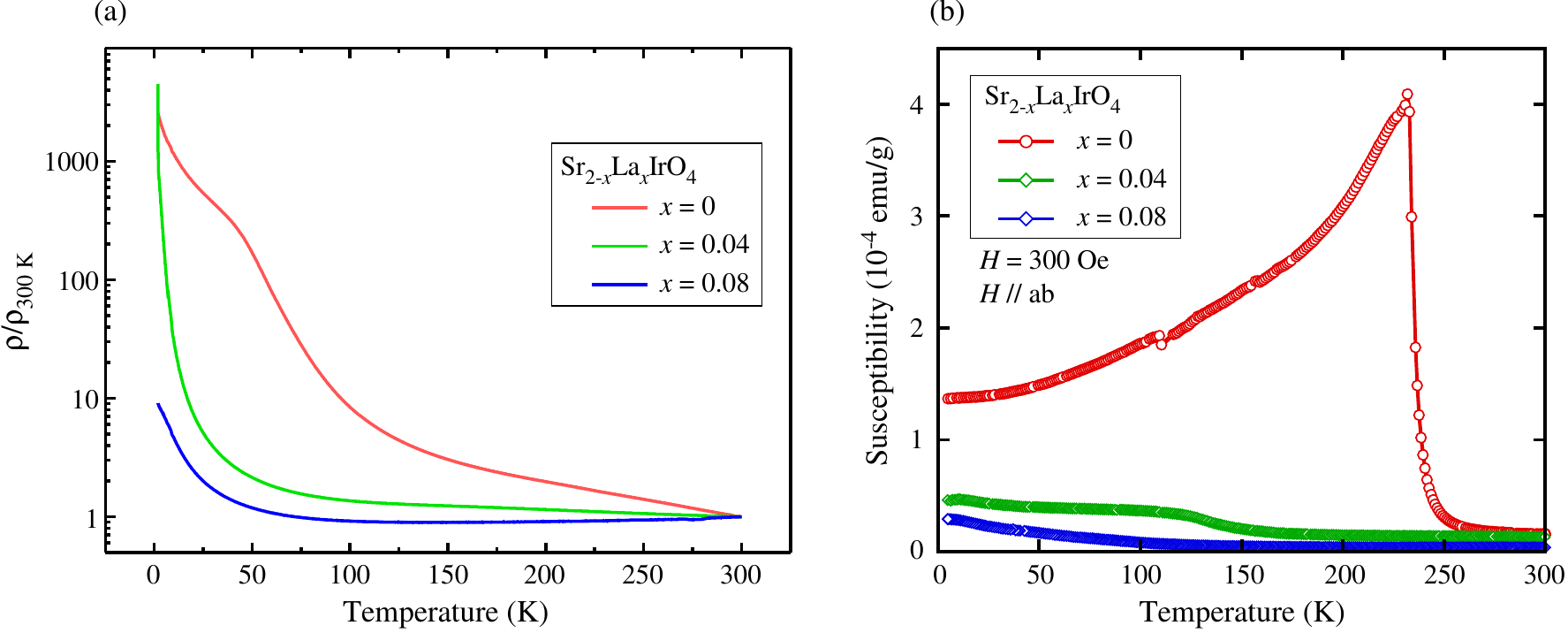}%
\caption{Temperature dependence of the resistivity (a) and the magnetic susceptibility (b) for Sr$_{2-x}$La$_x$IrO$_4$ ({\it x} = 0, 0.04, 0.08).
}
\end{figure}

\vspace{0.5in}

\section{ARPES intensity plots along high symmetry line}
	Figure S2 shows ARPES data of Sr$_{2-x}$La$_x$IrO$_4$ ({\it x} = 0, 0.04, 0.08) taken along $\Gamma$M, MX, and $\Gamma$X direction of Brillouin zone.  These are the same data as in the main text but here they are shown in an expanded energy scale for focusing on the dispersion of in-gap state.  Green curves are the dispersion derived from fitting peak positions in second derivatives shown in (a)-(c).  For fitting, we used second-order polynomial for $\Gamma$M direction and cosine function with phase as free parameter for MX and $\Gamma$X direction.  Further near {\it E}$_{\rm F}$ data and their curvature plots \cite{curve} are shown in Fig. S3.  Fig. S4 shows curvature plots for possibly better visualization of bands displayed in Fig. 2 in the main text.

\begin{figure}[t]
\includegraphics[width=6.5in]{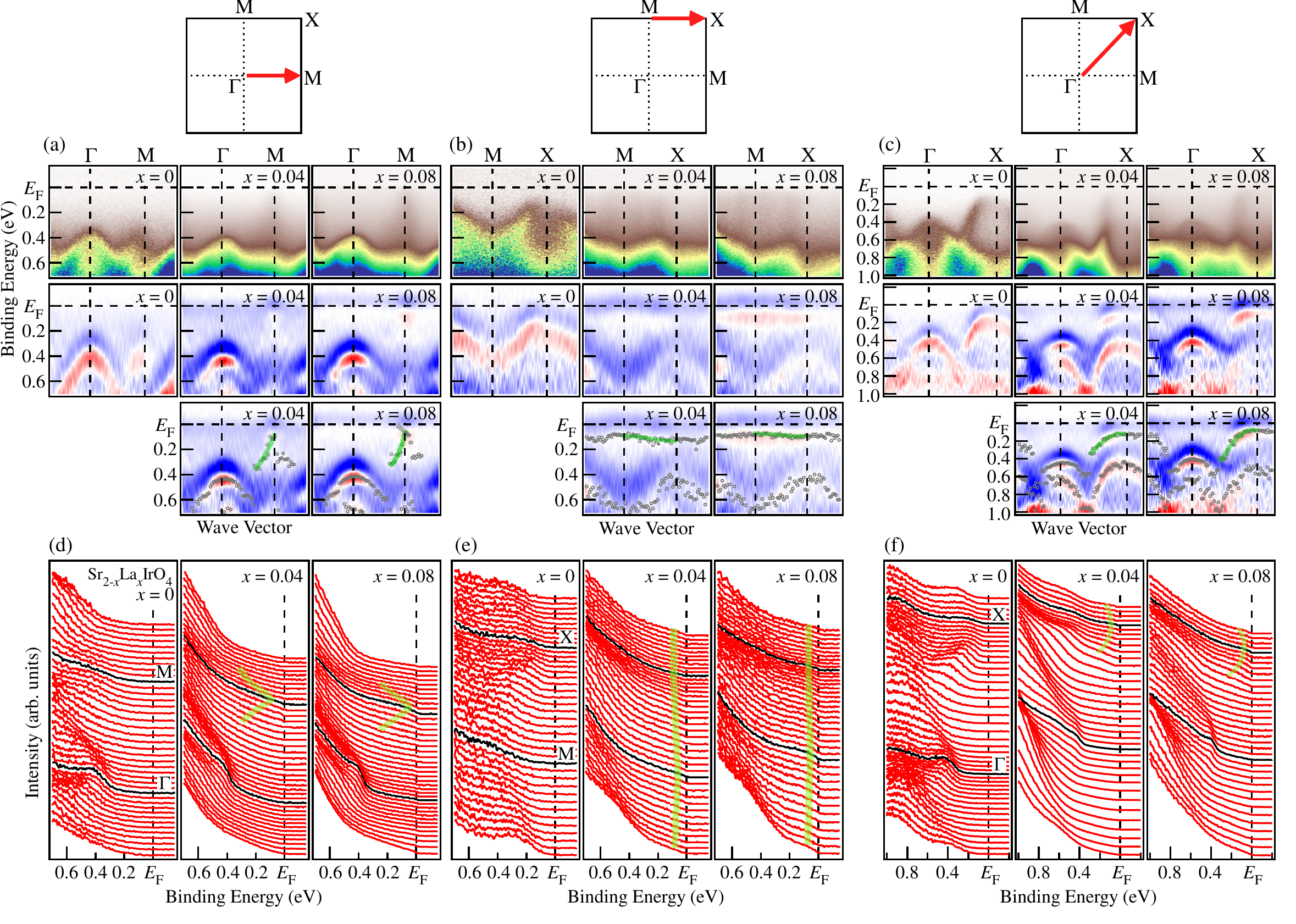}%
\caption{ARPES data of Sr$_{2-x}$La$_x$IrO$_4$ ({\it x} = 0, 0.04, 0.08) taken along $\Gamma$M, MX, and $\Gamma$X direction of Brillouin zone.  Top panel of (a)-(c) shows the intensity plots, middle panel shows their second derivatives, bottom panel shows second derivatives with peak positions of them (gray circles) and deduced dispersion of in-gap state (green curves). Gray line in x = 0.08 data ($\Gamma$M) are from Fig. S3. (d)-(f) Corresponding EDCs.  Green curves are the dispersion derived in (a)-(c).
}
\end{figure}

\begin{figure}[H]
\centering
\includegraphics[width=4in]{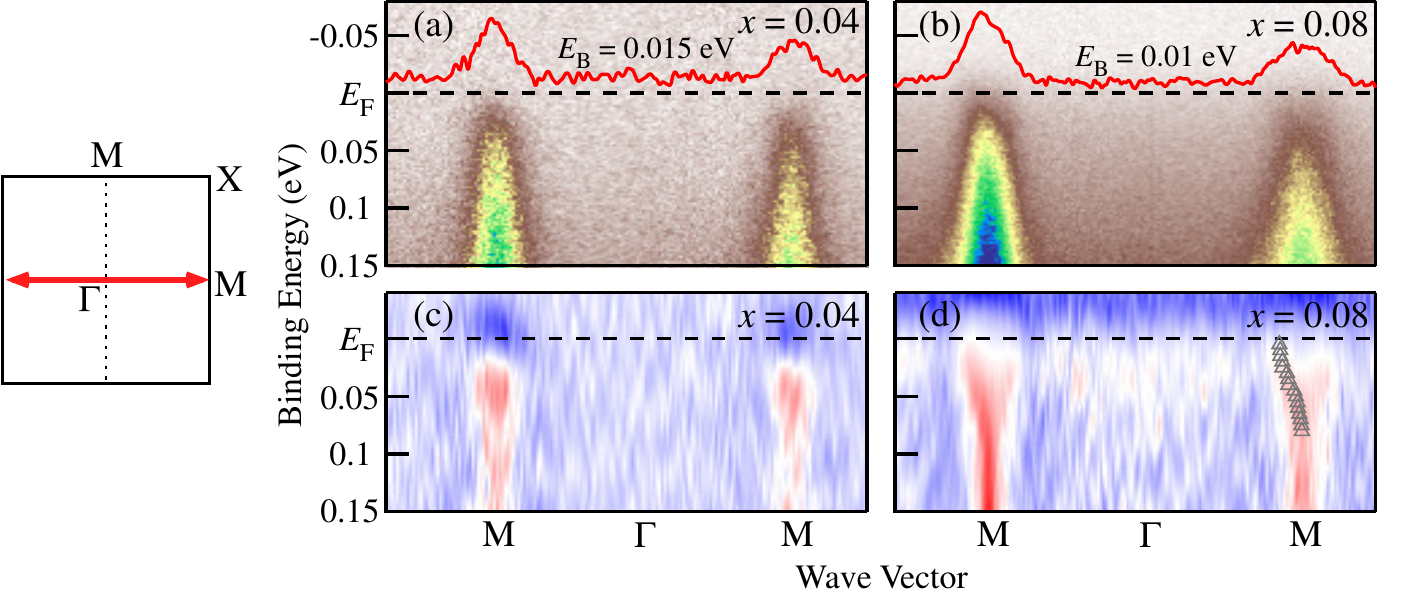}%
\caption{Near {\it E}$_{\rm F}$ ARPES data of Sr$_{2-x}$La$_x$IrO$_4$ ({\it x} = 0.04 (a), 0.08 (b)) along $\Gamma$M and their curvature plots ((c) and (d)).  MDCs are also shown.  Triangles are peak positions at each energy.
}
\end{figure}

\begin{figure}[h]
\includegraphics[width=5.2in]{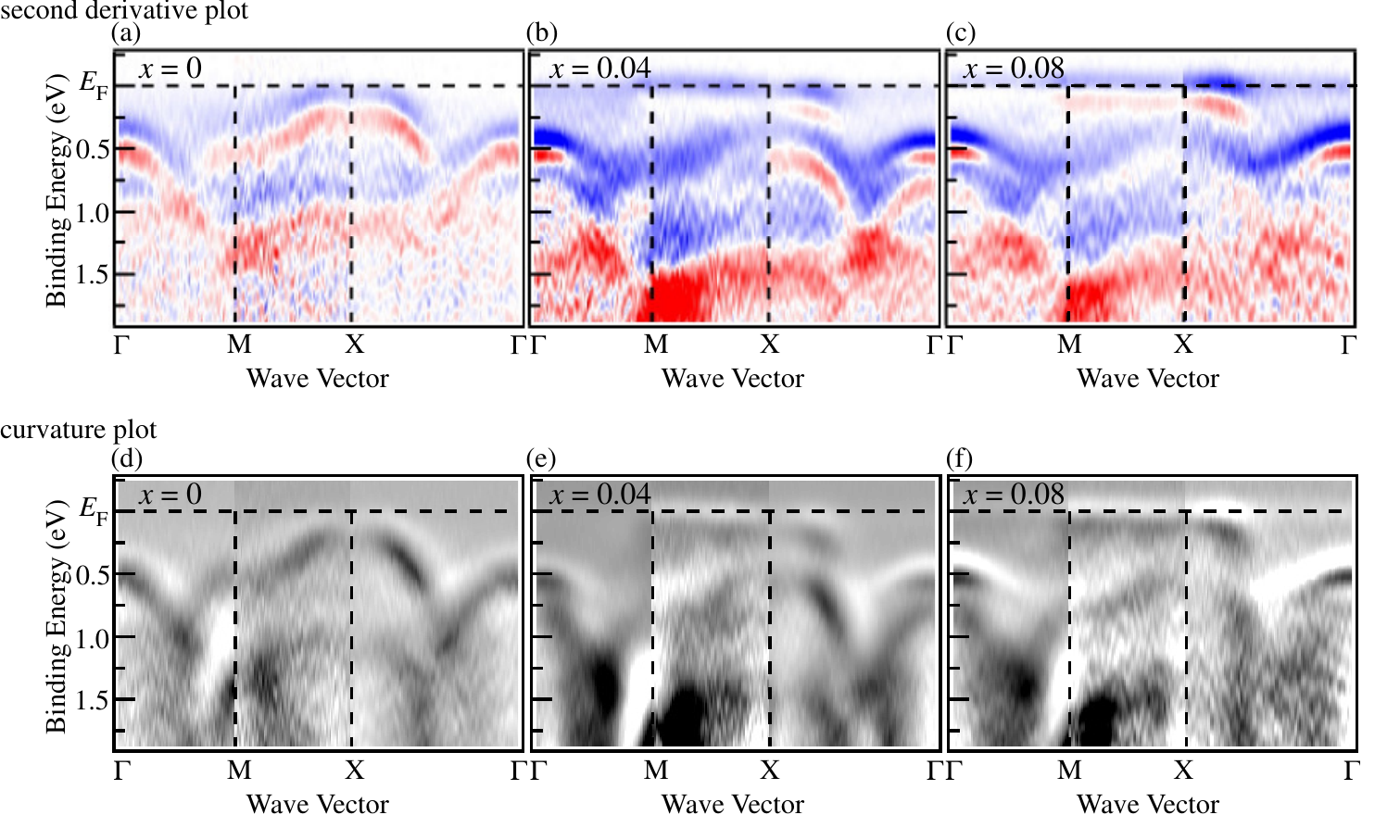}%
\caption{(a)-(c) Second derivative of ARPES spectra of Sr$_{2-x}$La$_x$IrO$_4$, same data as in Fig. 2 of main text.  (d)-(f) Curvature plot of the same data.
}
\end{figure}

\vspace{0.3in}
\newpage

\section{Comparison of dispersion between observed in-gap state and tight-binding model derived from LDA+SO ({\it U} = 0)}
	Figure S5 shows ARPES data of Sr$_{2-x}$La$_x$IrO$_4$ ({\it x} = 0, 0.04, 0.08) along $\Gamma$MX$\Gamma$ direction, which is the same data as in Fig. 2 in the main text.  Overlapped gray curves are calculated dispersion\cite{Delatorre} of tight-binding model with spin-orbit coupling ({\it U} = 0).  Among them, the band in which {\it J}$_{eff}$ = 1/2 is dominant is emphasized by thicker curve.  While the shape of remnant Fermi surface corresponds well to that of tight-binding {\it J}$_{eff}$ = 1/2 band, the energy position of in-gap state does not necessarily match with tight-binding {\it J}$_{eff}$ = 1/2 band.  The difference between them becomes the largest around X point where the size of excitation gap becomes the maximum, indicating that the band is shifted downward by the presence of the anisotropic gap.  We note that such an anisotropic energy-shift of band shows close resemblance to what has been reported in the pseudogap state in underdoped cuprate superconductor\cite{Marshall}.

\begin{figure}[h]
\includegraphics[width=6.5in]{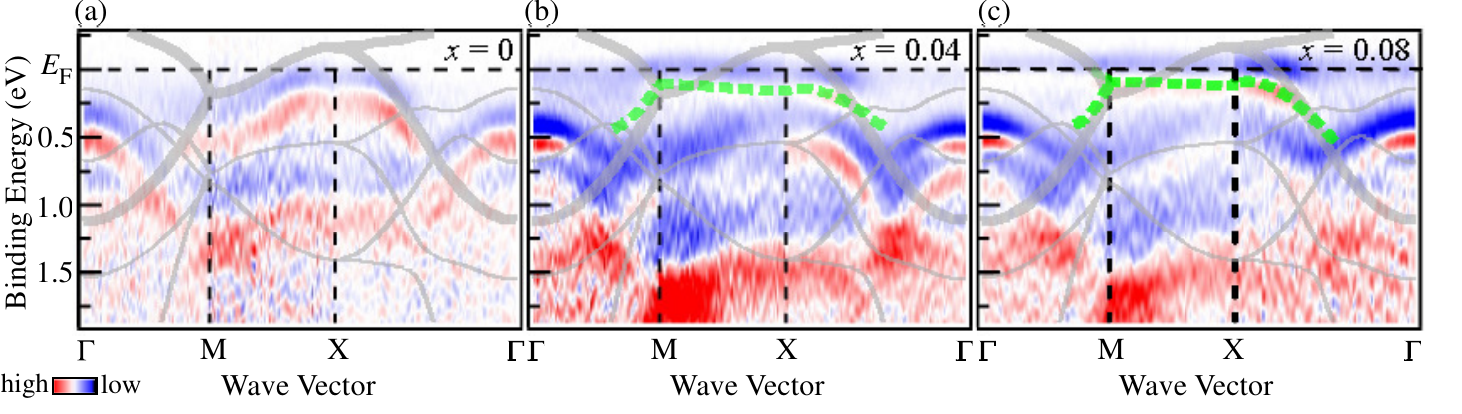}%
\caption{Second derivative plot of ARPES intensity in Sr$_{2-x}$La$_x$IrO$_4$ ({\it x} = 0, 0.04, 0.08) along $\Gamma$MX$\Gamma$ direction.  Calculated dispersion based on tight-binding model with spin-orbit coupling ({\it U} = 0)\cite{Delatorre} are shown as gray curves.  {\it J}$_{eff}$ = 1/2 band is shown as a thick curve.  Green curve shows dispersion of in-gap state deduced by peak positions in second derivative plot as shown in Fig. S2.
}
\end{figure}


\begin{thebibliography}{33}
\bibitem{TNnote}
	The simple phase separation is unlikely based also on macroscopic property in Fig. S1.  The magnetic susceptibility of La-doped samples with different doping level show behavior of magnetic order at different temperatures each other.
\end{thebibliography}
\end{document}